\documentclass[%
 preprint,
 amsmath,amssymb,
 aps,
]{revtex4-2} 
\usepackage{svg}
\usepackage{graphicx}
\usepackage{dcolumn}
\usepackage{bm}
\usepackage{float}
\usepackage{xcolor}
\usepackage{multirow}
\usepackage{tabularray}
\usepackage[colorlinks=true, urlcolor=blue, linkcolor=red]{hyperref}

\graphicspath{{./Figures_main/}}
\newcommand{\Wi}{\textrm{Wi}}
\newcommand{\trC}{\textrm{tr}\mathbf{C}}

\newcommand{\Conf}{\mathbf{C}}
\renewcommand{\u}{\mathbf{u}}
\newcommand{\Id}{\mathbf{I}}
\newcommand{\dgrad}{\cdot\nabla}
\newcommand{\taup}{\boldsymbol{\tau}}

\begin{document}


\title{A Period-Doubling Route to Chaos in Viscoelastic Kolmogorov Flow}

\author{Jeffrey Nichols}
\affiliation{%
 Department of Mathematics, University of California, Davis, Davis, CA 95616
}%
\author{Robert D.\ Guy}
\affiliation{%
 Department of Mathematics, University of California, Davis, Davis, CA 95616
}%
\author{Becca Thomases}
\affiliation{%
 Department of Mathematical Sciences, Smith College, Northampton, MA, 01063
}%

\date{\today}

\begin{abstract}  
Polymer solutions can develop chaotic flows, even at low inertia. This purely elastic turbulence is well studied, but little is known about the transition to chaos. In 2D channel flow and parallel shear flow, traveling wave solutions involving coherent structures 
are present for sufficiently large fluid elasticity. We numerically study 2D periodic parallel shear flow in viscoelastic fluids and show that these traveling waves become oscillatory and undergo a series of period-doubling bifurcations en-route to chaos.

\end{abstract}

\maketitle

\section{Introduction}\label{intro}

Polymer molecules in Newtonian solvents can align and stretch with the flow causing complex flow dynamics even at low or vanishing Reynolds number. This phenomena, referred to as  elastic turbulence, has been documented in experiments \cite{Steinberg2000,groisman2001efficient,ArratiaChannel2013} and seen in numerical simulations \cite{Berti2008PREArticle,lellep2024purely}, but there is no understanding of the mechanisms involved in the transition to turbulence. 
 Nevertheless, applications of elastic turbulence and viscoelastic instabilities abound including industrial polymer processing \cite{DennExtrusion, SangronizPolymersAndRheology}, 3D printing \cite{Tabakova3DPrintingPolymer, Siacor3DPrintingViscoelastic}, and enhancement of mixing in microfluidic devices \cite{ZhouMicrofluidicsReview, SousaMicroFluidicCrossSlot, ZizzariMixing}. 
 
Significant effort has gone into understanding the theoretical underpinnings of the flow instabilities and dynamics in elastic turbulence both with and without inertia \cite{Fouxen2001PRE, Fouxen2003_PhysicsOfFluids, MorozovIntroEssay,PRF2022Perspectives, Dubief_EIT_Review, KerswellMultiStabl2024}.  
Purely elastic turbulence (inertialess viscoleastic flow) was first identified by Groisman and Steinberg \cite{Steinberg2000} and relevant theory  \cite{Fouxen2001PRE, Fouxen2003_PhysicsOfFluids} and numerical studies have been performed to study the transitions in and properties of these flows \cite{Berti2008PREArticle, BertiKolmo2010, Becca2009PRLTransitionToMixOscil, Becca2011TransOscilMixing}.
While early efforts to understand viscoelastic flow instabilities focused on curved geometries \cite{shaqfeh1996purely,mckinley1996rheological},  more recent effort has  determined that chaotic flows can be reached through subcritical bifurcations and sustained in parallel shear flows \cite{Morozov2005PlaneCouette, MorozovIntroEssay, Jovanovic2010,ArratiaChannel2013, MorozovCoherentStructures,lellep2024purely}.

Coherent structures arising from the center-mode instability \cite{garg2018viscoelastic,khalid2021centre} in the elasto-inertial regime in channel flow were  first referred to as arrowheads \cite{page2020exact,DubiefFirstCoherentStructureEIT}.  They have subsequently been referred to as narwhals in the low Reynolds number regime \cite{lellep2023linear,MorozovCoherentStructures}. Previously these traveling wave structures (TWS) were identified (though unnamed) in Kolmogorov flow (periodic shear flow) at low inertia \cite{BertiKolmo2010}. In 2D Kolmogorov flow, it was observed that upon increasing the Weissenberg number, $\Wi$, these TWS lose stability and exhibit oscillations. At higher $\Wi$ the coherent structures repeatedly appear and disappear in chaotic flow \cite{BertiKolmo2010}. 
 Recently Lewy {\it et al.} \cite{Lewy2024} has shown that the linear instability observed in low Reynolds number Kolmogorov flow (\cite{BoffettaKolmogorov2005}) is the same instability that was found in pipe and channel flow and referred to as the center-mode instability \cite{garg2018viscoelastic}.

In 2D channel flow, also at low inertia,  these TWS co-exist with the stable uniform state and
can be obtained using finite size (not infinitesimal) perturbations at sufficiently high $\Wi$, but have not been directly associated to a turbulent regime \cite{MorozovCoherentStructures}.  In \cite{Buza_FiniteAmplitudeElasticWaves} it is noted that the domain length likely plays a role in the transition to turbulence in the presence of these TWS.
With non-negligible inertia in channel flow, TWS and chaotic regimes are found and can coexist \cite{DubiefFirstCoherentStructureEIT}.
While coherent TWS have been found in both Kolmogorov and channel flow, their relation to chaotic flow regimes is still not well understood \cite{Buza_FiniteAmplitudeElasticWaves, PRF2022Perspectives, kerswell2023asymptotics, KerswellMultiStabl2024,Lewy2024}.

Here we connect the TWS to chaos by examining zero Reynolds number, viscoelastic Kolmogorov flow in 2D, one of the simplest ways to numerically realize elastic turbulence. The periodic shear flow provides a framework to investigate transitions from coherent structures to turbulence. Using low spatial frequency forcing gives a single traveling wave solution, or narwhal, that as we increase $\Wi,$ progresses through a series of bifurcations, including period doubling bifurcations, and leads to the chaotic dynamics of elastic turbulence. 
This study uncovers a new set of bifurcations from coherent TWS to chaos in viscoelastic flow.  The simplicity of the system will  be useful in further analysis and exploration of this route to chaos. 

\section{Model and Methodology}\label{model}
To explore the transition to chaos we study the Kolmogorov flow in a 2D periodic domain. The system has been shown to exhibit chaos in the low Reynolds number regime \cite{BertiKolmo2010}. The viscoelasticity model used is the Stokes-Oldroyd B model, whose governing equations for the fluid velocity, $\u$, and the polymer stress tensor $\taup$ are as follows:
\begin{align}
&-\nabla p+\eta_s\Delta\u +\nabla\cdot\taup+\mathbf{F}=0,\\&\;\;\nabla\cdot\u=0,\\
&\tau+\lambda\stackrel{\nabla}\tau = \eta_p\left(\nabla\u+\nabla\u^T\right),
\end{align} where the upper-convected derivative is defined $\stackrel{\nabla}\taup \equiv \partial_t\taup+\u\dgrad\taup-\left(\nabla\cdot\u\taup+\taup\cdot\nabla\u^{T}\right).$  In our numerical simulations we evolve the conformation tensor $\Conf$, related by $\taup=\frac{\eta_p}{\lambda}(\Conf-\Id),$ and add polymer diffusion for numerical stability resulting in the evolution equation $\stackrel{\nabla}\Conf=-\frac{1}{\lambda}(\Conf-\Id)+\nu\Delta\Conf$ ($\nu=5\cdot 10^{-4}$).  The size of diffusion used here is within an order of magnitude of values considered realistically arising from diffusion of polymer center of mass in kinetic theory \cite{MorozovCoherentStructures}, and similar results at lower diffusion are given in Appendix \ref{A:diffusion}.  The parameter $\lambda$ is the polymer relaxation time, and $\eta_p,$ and $\eta_s$ are the polymer and solvent viscosities. Throughout this study, we fix $\eta_p/\eta_s=1/2$ and vary $\lambda$. 

Here we simulate Kolmogorov flow, \begin{equation}\label{anavel}\mathbf{u}=(-B\cos 4y,0),\end{equation} by prescribing a background driving force
 $\mathbf{F} = (-A\cos 4y,0)$ on the periodic domain  $[0 ,2\pi]\times [0, \pi/2].$ This domain was chosen specifically to support a single TWS.  Other domains that allow for interactions of multiple TWS are worthy of studying but beyond the scope of this manuscript. 
 
 This flow has an analytic solution for the conformation tensor (see Appendix \ref{A:analyticSoln}) and the amplitude $A$ is chosen such that the analytic solution has a maximum velocity of 4 in alignment with related works \cite{Berti2008PREArticle}. We define the Weissenberg number (non-dimensional elasticity parameter)  as $\Wi = \textrm{max}|\nabla U|\lambda ,$ where $U$ is the analytic velocity, and with the given analytic velocity, we have the relation $\Wi=16\lambda$.

A discretized Fourier pseudo-spectral method is used to solve the equations. The initial conditions for the conformation tensor are created from adding small spatially random perturbations to either the analytic solution or the final state of the solution from a different $\Wi$.

Results of the simulations are presented for the  strain energy ($\mathcal{E}$) and kinetic energy  ($\mathcal{K}$), \begin{equation}\mathcal{E}=\frac{1}{2\Wi}\int\int\trC \;dxdy,\;\; \mathcal{K}=\frac{1}{2}\int\int|\u|^2 \;dxdy.\label{eedef}\end{equation} Traveling wave solutions (TWS) are considered to be at steady state when initial transients have decayed and deviations from the new mean are less than $10^{-4}$. Similarly, periodic solutions must satisfy a convergence tolerance criteria over at least 50 periods (see Appendix \ref{A:criteria}). Most chaotic solutions were run for at least $3000$ time units.
More details of the numerical method and continuation are given in the Appendix \ref{A:methods}.

\section{Results}\label{results}

\begin{figure}
\includegraphics[width=\textwidth]{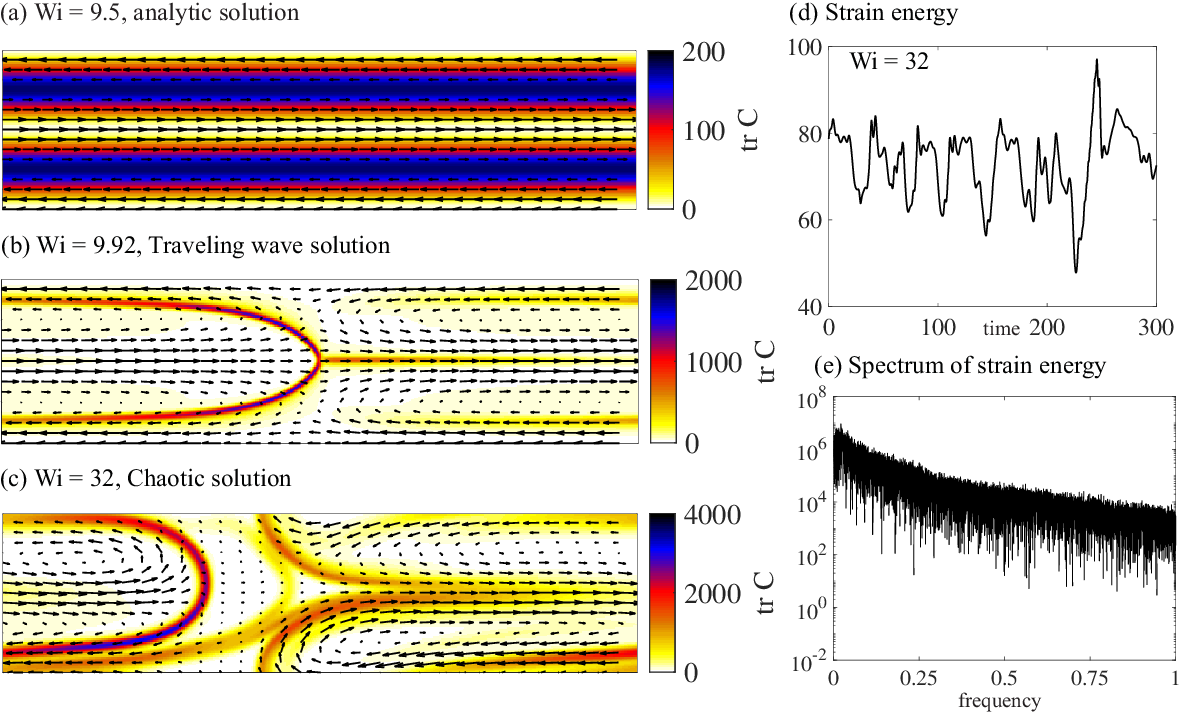}
\caption{$\trC$ with velocity vectors overlayed for (a) analytic solution at  $\Wi=9.5,$  (b) traveling wave solution at $\Wi=9.92$, (c) chaotic solution at $\Wi=32.$ (d) Time series of the strain energy for $\Wi=32$ over a representative time interval. (e) Spectrum of strain energy, $\Wi=32$. \label{fig:anatochaos}  }
\end{figure}

\begin{figure*}[ht]
  \includegraphics[width=.98\textwidth]{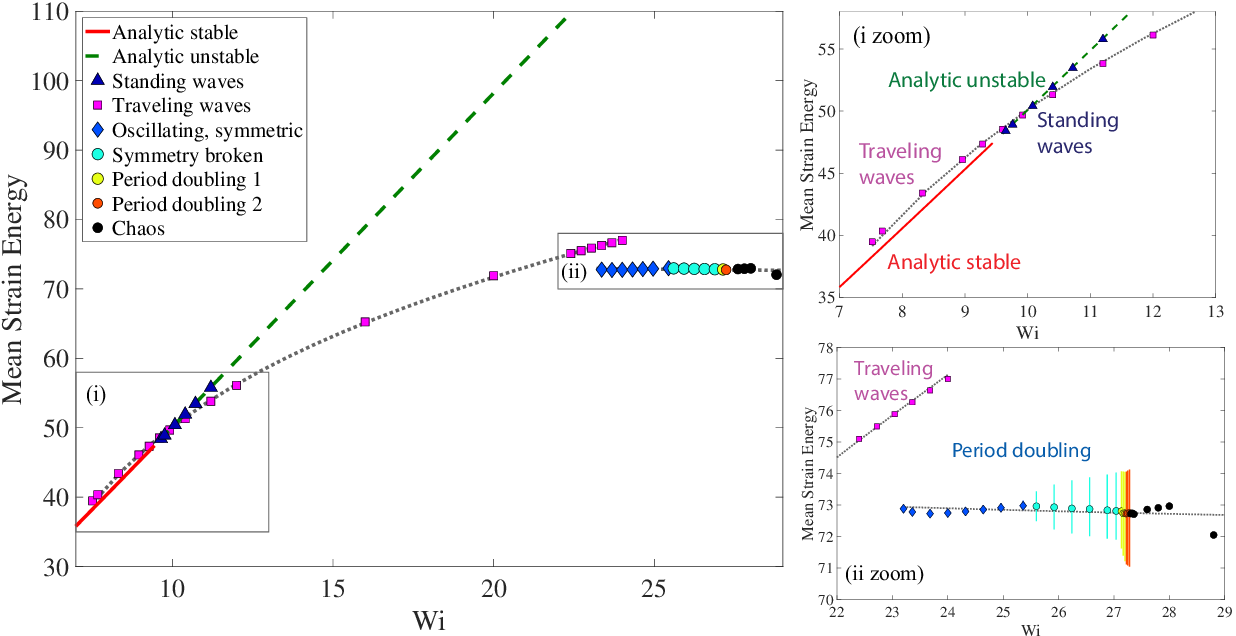}
  \caption{Mean strain energy and classification  of solution types as  a function of  $\Wi$. The analytic solution is linear in $\Wi$. A dotted line shows a square-root fit to the data for the TWS branch. A dotted line shows a constant fit to the data for the period doubling and chaotic branch. Regions highlighted in (i) and (ii) are expanded in (i zoom) and (ii zoom). The lines on the markers in (ii zoom) show the maximum and minimum values of the strain energy for the oscillating solutions.} \label{fig:bifdiag} 
\end{figure*}

We perform linear stability analysis (see Appendix \ref{A:LSAmethods}) and determine that the analytic solution is stable for $\Wi< 9.61.$ In the stable region the analytic solution for the $\trC$ has regions of high stress that correspond to where the shear rate is strongest; see  Fig.~\ref{fig:anatochaos}(a) for results at $\Wi=9.5$. Just above this threshold ($9.65\le\Wi\le 11.2$) small perturbations to the analytical solution result in standing wave solutions (see Appendix \ref{A:standingwaves}).  Above the threshold where the standing wave solutions exist, TWS are obtained starting from small perturbations from the analytic solution. By downward continuation the TWS are found to co-exist with the standing wave solutions and the analytic solutions.  As an example and to illustrate the TWS structure, $\trC$ and velocity for the TWS at $\Wi=9.92$ are shown in Fig.~\ref{fig:anatochaos}(b).

In Kolmogorov flow there is a transition to a chaotic state at higher $\Wi$ \cite{BertiKolmo2010}. 
We similarly see chaotic behavior for $\Wi\gtrsim  27.3.$ A time snapshot of $\trC$  and flow 
 are  given in Fig.~\ref{fig:anatochaos}(c) for $\Wi=32,$  where a structure resembling the narwhal is present. Movies  of the dynamics showing intermittent formation and destruction of the narwhal structure are given in Appendix \ref{A:movies}.  In Fig.~\ref{fig:anatochaos}(d) we see aperiodic behavior of the  strain energy over time,  and the energy spectra, Fig.~\ref{fig:anatochaos}(e), shows a wide range of excited frequencies.  This behavior is typically classified as elastic turbulence \cite{Berti2008PREArticle}.

 Interestingly, the  analytic solution becomes linearly stable again for $\Wi\gtrsim 36.9.$ In this high $\Wi$ regime there is still a nonlinear route to chaos via numerical continuation from chaotic solutions. More details of the linear stability analysis are given in Appendix \ref{A:LSAmethods}.

In what follows we return to the narwhal TWS and consider what happens as we systematically vary $\Wi.$ We obtain oscillatory solutions that exhibit a set of period doubling bifurcations leading  up to the chaotic regime. In Fig.~\ref{fig:bifdiag} we display the temporal mean of the strain energy $\mathcal{E}$ as a function of $\Wi$ and categorize the solution types. 
As we decrease $\Wi$ below 9.92, the TWS  persists down  to $\Wi=7.52$ (see Fig.~\ref{fig:bifdiag}(i)).  Below this value, the analytic solution is the only solution  we found by continuation. In  the range $7.52<\Wi<9.61$ both the traveling wave and analytic solutions are stable. This is similar to channel flow \cite{MorozovCoherentStructures, KerswellMultiStabl2024} where the stable traveling wave solution and analytic solutions co-exist above some threshold in $\Wi$, but in channel flow the base state remains linearly stable.

Increasing $\Wi$, the numerical continuation produces steady traveling narwhal solutions until about $\Wi\approx 24,$ when a new lower energy stable branch appears as a subcritical bifurcation.  A bifurcation diagram with kinetic energy is given in Appendix \ref{A:KEbifdiag} and shows a more significant drop in $\mathcal{K}$ for the TWS branch.  Both steady TWS and oscillating solutions exist for $23\lesssim \Wi\lesssim 24,$  however at $\Wi=22.88$ a downward numerical continuation from the oscillating solution returns to the steady traveling wave solution. 

\begin{figure}
\includegraphics[width=.7\textwidth]{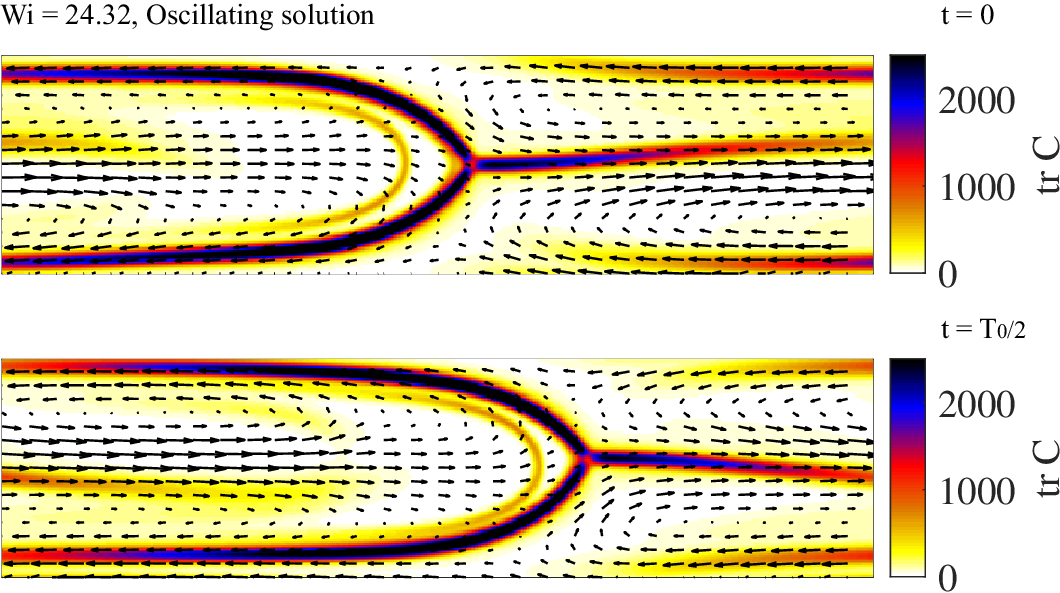}
\caption{$\trC$ and velocity vectors  for $\Wi = 24.32$ at $t=0$ and $t=T_0/2,$ period $T_0=6.17.$ }\label{fig:oscillations} 
\end{figure}

In Fig.~\ref{fig:oscillations} we show $\trC$ in the oscillatory regime where the narwhal ``tusk" exhibits a periodic vertical deformation; see Appendix \ref{A:movies} for movies. The first oscillatory solutions that emerge after traveling waves have a vertical symmetry which is lost as $\Wi$ increases; see Appendix \ref{A:symmm} for more details.

The strain energy for the analytic solutions demonstrates a linear scaling in $\Wi$ which is seen in Fig.~\ref{fig:bifdiag}, but the  strain energy fits a 1/2 power-law scaling in the TWS regime.
 A sub-critical bifurcation to oscillating symmetric solutions appears as another lower energy branch with approximately constant(dotted line) scaling with $\Wi.$
As $\Wi$ increases the amplitude of the oscillations are initially small but grow  with $\Wi$ as indicated by vertical bars in Fig.~\ref{fig:bifdiag}(ii).

\begin{figure}
\includegraphics[width=\textwidth]{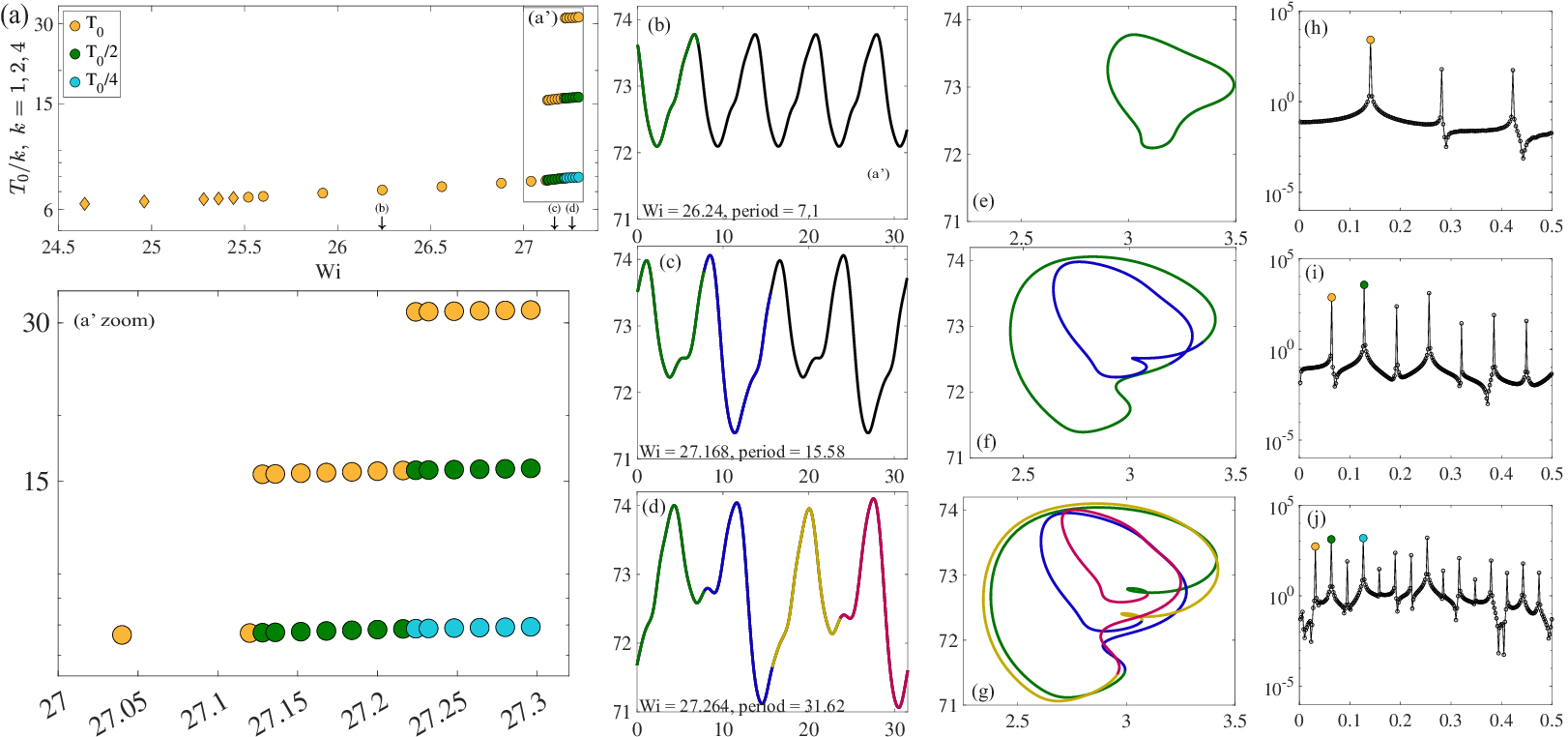}
\caption{(a) Period and harmonics as a function of $\Wi$. Diamond markers represent symmetric solutions.
(b)-(d) Strain energy over time for $\Wi=26.24, 27.168, 27.264$ with a period highlighted in color(s). (e)-(g) Strain energy versus kinetic energy over a period with color labeling matching (b)-(d). (h-j) Spectrum of strain energy with dominant frequencies highlighted, corresponding to the period and harmonics labeled in (a).)  \label{fig:cascade}  }
\end{figure}

In Fig.~\ref{fig:cascade}(a) the period of the strain energy,  $T_0,$ 
is plotted for a portion of the branch of periodic solutions below the chaotic regime, $24.5<\Wi<27.3.$  To identify the period we use the Fourier transform of the strain energy of solutions; see for example Figs.~\ref{fig:cascade}(h)-(j).
The period slowly increases in $\Wi,$ and near $\Wi=27.12$ the  period jumps from  $7.72$ to $15.47$.
The period again doubles around  $\Wi= 27.216.$ The harmonics $T_0/2$ and $T_0/4$ are highlighted for the first and second doublings to help illustrate the period doubling phenomena. 

We further illustrate period doubling by visualizing the strain energy (Figs.~\ref{fig:cascade}(b)-(d) ) and its spectrum (Figs.~\ref{fig:cascade}(h)-(j)) for three solutions on the period doubling branch corresponding to $\Wi = 26.24$, 27.168, and 27.264.

For each of these three $\Wi$ we additionally visualize the strain energy versus the kinetic energy, in Figs.~\ref{fig:cascade}(e)-(g). For solutions with the shortest period this is a loop, and at each successive doubling the loop splits in two.  The colors of the loops correspond to the  strain energy over time.

\begin{figure}
\includegraphics[width=\textwidth]{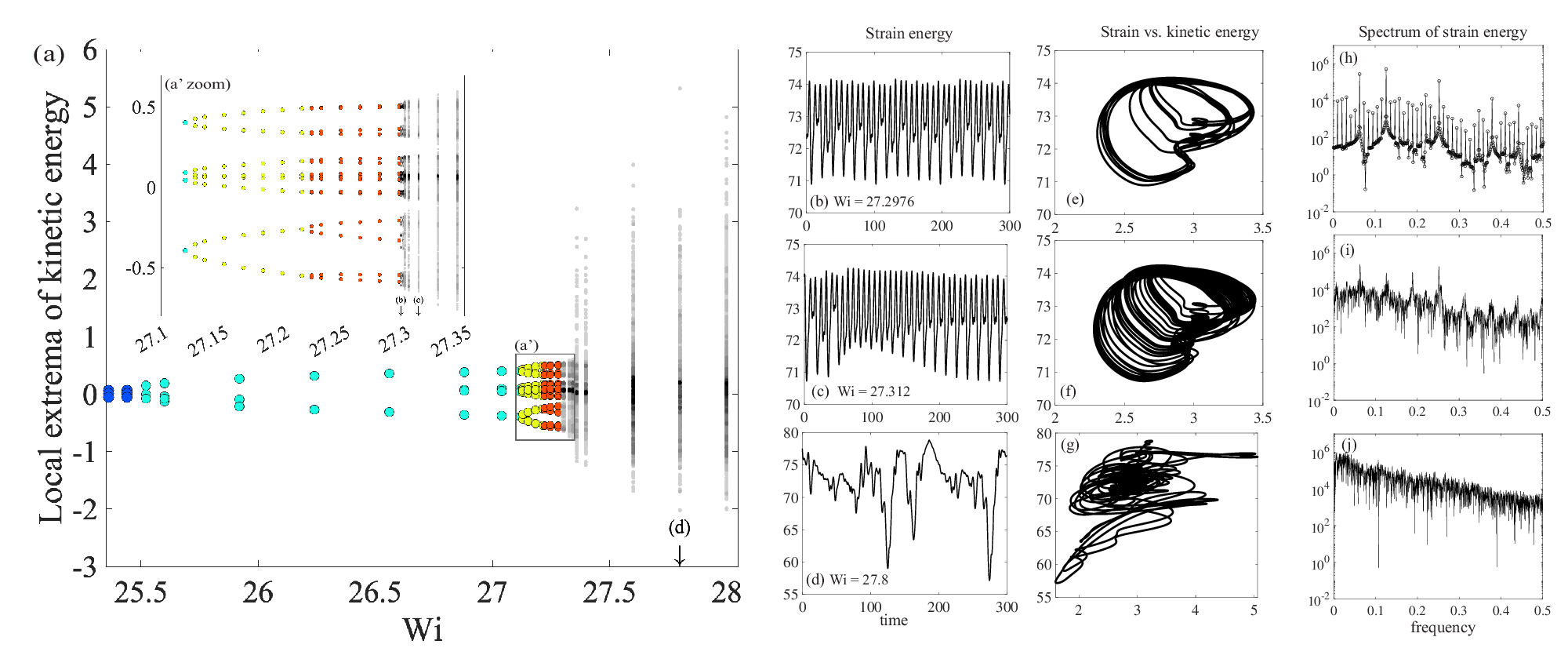}
\caption{\label{fig:chaos} (a) The local extrema of the deviation from the mean kinetic energy. Periodic solutions are colored according to the legend in the bifurcation diagram (Fig.~\ref{fig:bifdiag}). (b)-(d) Strain energy over time for $\Wi=27.2976, 27.312, 27.8$ over  a representative time interval. (e)-(g) Strain energy versus kinetic energy. (h)-(j) Spectrum of strain energy.}
\end{figure}

In Fig.~\ref{fig:chaos}(a), the bifurcations in the region $25<\Wi<28.1$  are visualized using local extrema of  deviations from the mean of the kinetic energy.  For solutions below $\Wi\le 27.296$ extrema of the periodic solutions are colored according to the legend in the bifurcation diagram of Fig.~\ref{fig:bifdiag}. The grayscale markers for $\Wi> 27.296$ also represent local extrema, but in this regime the flow has become increasingly chaotic and the extrema no longer repeat periodically. In this regime the extrema are binned according to the number of extrema with similar values in that bin with darker markers indicating more occurrences in a bin. This branching diagram has similarities with orbit diagrams of iterated maps \cite{strogatz2018nonlinear}.   Below the first period doubling ($\Wi\le 27.12$) there are between 2-4 local extrema for each solution. At the first doubling (cyan to yellow markers) 4 local extrema become 8, and at the next doubling (between yellow and orange markers) the 8 extrema split into 16.

Beyond $\Wi\approx 27.3$  the flow in these regions may have very long periods, may be intermittently chaotic, or may be chaotic.  For example, at $\Wi=27.2976$ (Figs.\ref{fig:chaos}(b,e,h)) the lowest frequency indicates a period of $95.$  
Increasing $\Wi$ to $27.312$ (Figs.\ref{fig:chaos}(c,f,i)) the oscillations in the strain energy show some structure but also exhibit periods of intermittency. The frequency still shows distinct peaks in the signal, but there is also a lot of additional noise with many frequencies being active. Increasing further to $\Wi=27.8$ (Figs.~\ref{fig:chaos}(d,g,j)) the dynamics more closely resemble those previously described as chaotic for $\Wi=32$ (Figs.~\ref{fig:anatochaos}(d,e)). A link for movies from all solution types is given in Appendix \ref{A:movies}.

\section{Conclusions}\label{conclusions}
Using forward simulations and numerical continuation we identified a subcritical bifurcation from traveling wave solutions to oscillating solutions in a 2D viscoelastic fluid driven by a periodic shear flow. The oscillating solutions then undergo a series of period-doubling bifurcations below the chaotic regime. As these bifurcations occur, waves in the narwhal ``tusk" appear to oscillate and interact with other regions of high stress in the coherent structure. This self-interaction of a single narwhal structure appears to lead to longer periods of oscillations as $\Wi$ increases until a chaotic regime is reached.  The specific bifurcation route may be dependent on this domain as only a single narwhal structure is found and the periodic boundary conditions introduce symmetries into the system.  Nevertheless, this bifurcation structure provides a novel systematic connection between coherent structures to chaotic dynamics in a viscoelastic fluid.  

Chaotic flows have been seen in Kolmogorov flow with higher frequency forcing \cite{Berti2008Book,BertiKolmo2010,Lewy2024} but the transition to chaos involves more complex dynamics due to the interaction of multiple coherent structures. However, the existence of the narwhal traveling wave solution does not necessarily imply that a flow will become chaotic; in fact a single structure is believed to be stable in channel flow \cite{MorozovCoherentStructures}, though there are indications that the stability of the structure may depend on channel length \cite{Buza_FiniteAmplitudeElasticWaves}.

Unlike 2D, in 3D the narwhal in a channel appears to become unstable leading to chaotic flows  \cite{MorozovLinearStab3DNarwhals, lellep2024purely}. The relatively simple framework of a single narwhal in 2D allows the identification of bifurcations that may also be underpinning those found in other more complex flows on the route to chaos.

The single narwhal structure is reminiscent of channel flow yet it still exhibits chaos; this comparatively simple flow and subsequent flow transitions give the first description of a mechanistic route to chaos.  Further study of this system may help unlock other mechanisms driving the transition to chaos in more complex systems through numerical simulations and analysis.

\nocite{HOU_Filter}

\appendix

\renewcommand{\thefigure}{A\arabic{figure}}
\renewcommand{\theequation}{A\arabic{equation}} 
\renewcommand{\thetable}{A\arabic{table}}

\section{Analytic Solution} \label{A:analyticSoln}

The system we solve is
\begin{align}
& -\nabla p+\Delta\u+\xi/\lambda\nabla\cdot\Conf+\mathbf{f}=0,\label{momentum eq}\\
&\nabla\cdot\u=0,\\
& \stackrel{\nabla}\Conf=-\frac{1}{\lambda}(\Conf-\Id)+\nu\Delta\Conf,\label{Conf eq}
 \end{align}  
 with $\mathbf{f} = (-A\cos ny,0).$ The upper-convected derivative is defined $\stackrel{\nabla}\Conf \equiv \partial_t\Conf+\u\dgrad\Conf-\left(\nabla\u\cdot\Conf+\Conf\cdot\nabla\u^{T}\right).$ Given the polymer and solvent viscosities, $\eta_p$ and $\eta_s$,  we define 
 the coupling constant $\xi=\frac{\eta_p}{\eta_s}.$ \\

For the given force the analytic solution for the velocity is
\begin{equation}\label{anavelsm}(u,v)=(-B\cos ny,0),\end{equation} and the conformation tensor is
\begin{align} \label{anaCsm}
    C_{11} &= 1 +  E\sin^2(ny)+2\lambda\nu En^2,\\
    C_{12} &= \frac{Bn\lambda}{1+\lambda\nu n^2}\sin(ny),\\
    C_{22} &= 1,
\end{align}  
where $ E = 2(Bn)^2\left(\frac{1}{\lambda^2}+\frac{5\nu n^2}{\lambda}+4\nu^2n^4\right)^{-1}$ and  $ A = Bn^2\left(1+\frac{\xi}{1+\nu n^2\lambda}\right).$  In the simulations we set $B=4,$  $n=4,$  $\xi=1/2,$  $\nu = 5\cdot10^{-4},$ and vary the relaxation time $\lambda.$ We define the Weissenberg number (non-dimensional elasticity parameter)  as $\Wi = \textrm{max}|\nabla U|\lambda ,$ where $U$ is the analytic velocity, and with the given analytic velocity, we have the relation $\Wi=16\lambda$. 

\section{Methods}\label{A:methods}

\subsection{Numerical Solutions} \label{A:numdetails}

We perform direct numerical simulation of the system given in Eqs.~\eqref{momentum eq}-\eqref{Conf eq} on a $[0,2\pi]\times[0,\pi/2]$ domain with doubly periodic boundary conditions.  To advance in time from $t$ to $t+\Delta t$, we first determine $\Conf(t+\Delta t)$ using $\Conf(t)$ and $\u(t)$.  This step is done with a splitting method, using Runge-Kutta 4 (RK4) to update all but the diffusive term, which is done with backward Euler. We use a psuedospectral method, with spatial derivatives computed in Fourier space, and nonlinear multiplication done in real space.  Prior to this multiplication a spectral filter,
\begin{equation}
    \zeta(k_x,k_y) = \exp\left[-36\left(\frac{k_x}{\max(k_x)}\right)^{36}\right]\boldsymbol{\cdot} \exp\left[-36\left(\frac{k_y}{\max(k_y)}\right)^{36}\right],
\end{equation} is applied to each factor \cite{HOU_Filter}.\\

 Once $\Conf(t+\Delta t)$ is known, we use it to determine $\u(t+\Delta t)$. To do this, we invert the Stokes equations in Fourier space.  We take the mean velocity to be 0, meaning we can invert the Laplacian in Fourier space as
\[ 
\widehat{\Delta^{-1} f} =  \begin{cases}
    0 \quad &|k|=0\\
    \frac{-1}{|k|^2}\hat{f} \quad &|k|>0.
\end{cases}
\]

 For the simulations presented in the main text, a $512\times128$ grid was used with a timestep of $\Delta t=1.25\cdot10^{-3}$.  For the simulations with lower diffusion (see Appendix \ref{A:LowDiffResults}), the grid was refined to $1024\times256$ and the time step was halved.
 
\subsection{Perturbations}\label{A:pert}
The initial condition used for each simulation in this study consists of either the analytic solution or, in the case of numerical continuations, the final state of a previous simulation.  In either case, a small random perturbation is applied to the first component of the conformation tensor, $C_{11}$.  The initial conformation tensor's first component is
\begin{equation*}C_{11,init} = C_{11} + \delta(x,y),\end{equation*}
where $\delta(x,y)$ is the perturbation, which is of the form
\begin{equation*}\delta(x,y) = 10^{-6}\cdot M\cdot |R(x,y)|.\end{equation*}
At each $(x,y)$ location, the value of $R$ is random, drawn from a standard normal distribution.  $M$ is the max of $C_{11}$ before the perturbation is applied.

\subsection{Criteria for Steady Oscillations}\label{A:criteria}
For simulations involving oscillatory narwhals, care must be taken to ensure that a steady state has been reached because transient dynamics, sometimes manifesting as quasi-periodic behavior, can exist for a very long time. To be considered steady oscillations, we impose requirements on the number of oscillations and the consistency of maximum and minimum values seen through all the oscillations.  To test a set of oscillations, we remove early data containing transient dynamics and apply the following procedure to the remaining data.  The procedure is applied separately to the set of minima and to the set of maxima in the data.

Let $x_i$ represent either an absolute minimum or maximum at a sample point in the data collected during one period of oscillation.  We fit a second degree polynomial to the sample points around this value, $\{x_{i-1},\, x_i,\, x_{i+1}\}$ and find the extremum of this polynomial, which we denote $p_i$.  This helps account for deviations due to discrete sample points not being aligned with the true location of an extreme value.

Now, define $P=\{p_1,\, p_2,\, ...,\, p_N\}$ to be the set of all such maximum or minimum (not both) values under consideration. Let $m$ be the mean of $P$ and let $A$ be the average amplitude of the oscillations.  We compute the following three averages, which reflect deviation from the mean at the start ($S$), end ($E$), and through all ($L$) of the signal, normalized by the amplitude:
\begin{align}
    S &= \frac{1}{A}\cdot\frac{1}{5}\cdot\sum_{i=1}^5|p_i-m|,\\
    E &= \frac{1}{A}\cdot\frac{1}{5}\cdot\sum_{i=N-4}^N|p_i-m|,\\
    L &= \frac{1}{A}\cdot\frac{1}{N}\cdot\sum_{i=1}^N|p_i-m|.
\end{align}
The requirements for steady oscillations are
\begin{itemize}
    \item $N \geq 50$
    \item $L<10^{-3}$
    \item $|S-L|<10^{-4}$
    \item $|E-L|<10^{-4}$
\end{itemize}

 The first requirement is simply that the data includes at least 50 oscillations. The requirement on $L$ ensures that across all the data under consideration, extrema occur at approximately the same value.  The requirements on $S$ and $E$ ensure that any transient dynamics early in the signal have sufficiently died out and that the end of the signal is not showing evidence of instability.

\subsection{Run-times for non-periodic data}\label{A:runtimes}
\begin{table}[h]
    \centering
    \begin{tabular}{|c|c|}
        \hline
        ${\Wi}$ & {Simulation Length}\\
        \hline
        27.2976 & 2780\\
        \hline
        27.2992 & 3200\\
         \hline
        27.3000 & 4000\\
         \hline
        27.3008 & 3520\\
         \hline
        27.3040 & 3000\\
        \hline
        27.3120 & 2500\\
        \hline
        27.3280 & 4000\\
        \hline
        27.3440 & 4000\\
        \hline
        27.3600 & 3510\\
        \hline
        27.4000 & 3000\\
        \hline
        27.6000 & 3000\\
        \hline
        27.8000 & 3000\\
        \hline
        28.0000 & 10030\\
         \hline
        28.8000 & 1000\\
         \hline
        30.4000 & 1000\\
         \hline
        32.0000 & 10040\\
         \hline
    \end{tabular}
    \caption{Simulation lengths (in time) for non-periodic data}
    \label{tab:simLengthsAperiodic}
\end{table}

$\Wi=27.2960$ is the highest value at which we obtained data with steady oscillations.  Above this value, we ran simulations to extended times to ascertain whether stable periodic solutions would emerge.

Just above $\Wi=27.2960$, simulations were run for at least 2500 time units. As a reference point, the 3 largest $\Wi$ simulations with steady oscillations had reached steady states before a time of 1600.  Two simulations ($\Wi=28$ and $\Wi=32$) were run for over 10,000 time units.  With these firmly established as aperiodic, the simulations with $28 < \Wi < 32$ were only run for 1000 time units, which was sufficient to generate the desired data.  All run times are listed in Table~\ref{tab:simLengthsAperiodic}.

\section{Linear Stability Analysis}

\subsection{Methods}\label{A:LSAmethods}

We begin by linearizing the Stokes-Oldroyd-B equations about the analytic solution, which we denote $(\Conf_0,\u_0, p_0).$ Adding a perturbation, we have $(\Conf,\u,p) = (\Conf_0+\tilde \Conf, \u_0+\tilde \u,p_0+\tilde p)$.  Plugging this into the governing equations and dropping quadratic terms yields an equation for $\tilde \Conf$, subject to a constraint equation involving $\tilde \u$ and $\tilde p$.  We can write this as
\begin{align}
    \partial_t\tilde \Conf & = N(\u_0,\tilde \Conf) + N(\tilde \u, \Conf_0) - \frac{1}{\lambda} \tilde \Conf +\nu \Delta \tilde \Conf, \label{conf_eqn} \\
    L(\tilde \u,\tilde p) &= \frac{-\xi}{\lambda}\nabla \cdot\tilde \Conf, \label{stokes}
\end{align}
where $N(\u,\Conf) = \u\cdot\nabla \Conf -\left(\nabla \u\cdot\Conf + \Conf\cdot\nabla \u^T\right)$ is the collection of nonlinear terms in the Stokes-Oldroyd-B equations and $L$ is the Stokes operator.  The velocity perturbation is determined by the perturbation to the conformation tensor: $\tilde\u=L^{-1}(\frac{-\xi}{\lambda}\nabla \cdot\tilde \Conf)$. Thus Eqs.~\eqref{conf_eqn}-\eqref{stokes} can be written in terms of $\tilde\Conf$ as $\partial_t \tilde \Conf=A(\Conf_0,\u_0,p_0)\tilde\Conf$ where $A$ is a linear operator depending on the steady solution.

We then build the discrete operator $A(\Conf_0,\u_0,p_0)$ directly.  That is, we assemble the operator column-by-column by applying the RHS of Eq.~\eqref{conf_eqn} to unit vectors.  Once the operator is built, the Matlab function `eig' is used to compute the eigenvalues of the operator.  The analytic solution is determined to be stable when all eigenvalues have negative real part, and unstable otherwise.

As $\Wi$ is increased, Kolmogorov flow first loses and later regains linear stability, and thus there are two critical values.  To determine each of these values we use a bisection search method.  Before this process begins, we must manually find a lower bound $\Wi_l$, and an upper bound $\Wi_u,$ so that the critical value lies between these bounds, i.e. $\Wi_l < \Wi_c < \Wi_u$.  Testing the midpoint $(\Wi_l + \Wi_u$)/2 for stability allows the search window to be cut in half.  This process is repeated until the upper and lower boundaries are within a tolerance of $10^{-3}$ of one another.  At this point, the average of the upper and lower boundaries is taken to be the critical value.  

\subsection{Grid Resolution}\label{A:grid}

When building the discretized operator $A(\Conf_0,\u_0,p_0)$, the grid resolution is limited by the large memory requirement of building the operator.  However, coarse resolutions yield surprisingly accurate results due to the low mode nature of the analytic solution.  Using a $96\times24$ grid for a $[0 ,2\pi]\times [0, \pi/2]$ domain was found to be satisfactory in terms of both accuracy and memory.  Higher resolutions were occasionally used to verify results.  For example, the data in Fig.~\ref{fig:diffusionAndWiC} was computed with $Ny=24$ and a subset verified with $Ny=32$.  To compare resolutions, consider Table~\ref{tab:Wi_c and Res} below, showing values found for $\Wi_c$ where stability is lost. A value of $\Wi_c=9.61$ was reported in the main text.  Time dependent simulations agree with these results; simulations at $\Wi=9.60$ and $\Wi=9.65$ were stable and unstable, respectively.

\begin{table}[h]
    \centering
    \begin{tabular}{|c|c|}
         \hline
        $Ny$ & $\Wi_c\; (\pm \, 5\cdot10^{-4}$)\\
         \hline
        16 & 9.6189\\
         \hline
        24 & 9.6151\\
         \hline
        32 & 9.6133\\
         \hline
        48 & 9.6136\\
         \hline
    \end{tabular}
    \caption{Critical $\Wi$ values found at various grid resolutions.  As $\Wi$ is increased past $\Wi_c$, Kolmogorov flow switches from linearly stable to linearly unstable.  This computation is done with the stress diffusion coefficient $\nu = 5\cdot10^{-4}$.}
    \label{tab:Wi_c and Res}
\end{table}

\subsection{Results}  \label{A:LinStabResults}

\section{Standing Waves} \label{A:standingwaves}
\begin{figure}[h]
    \centering
    \includegraphics[width=0.75\linewidth]{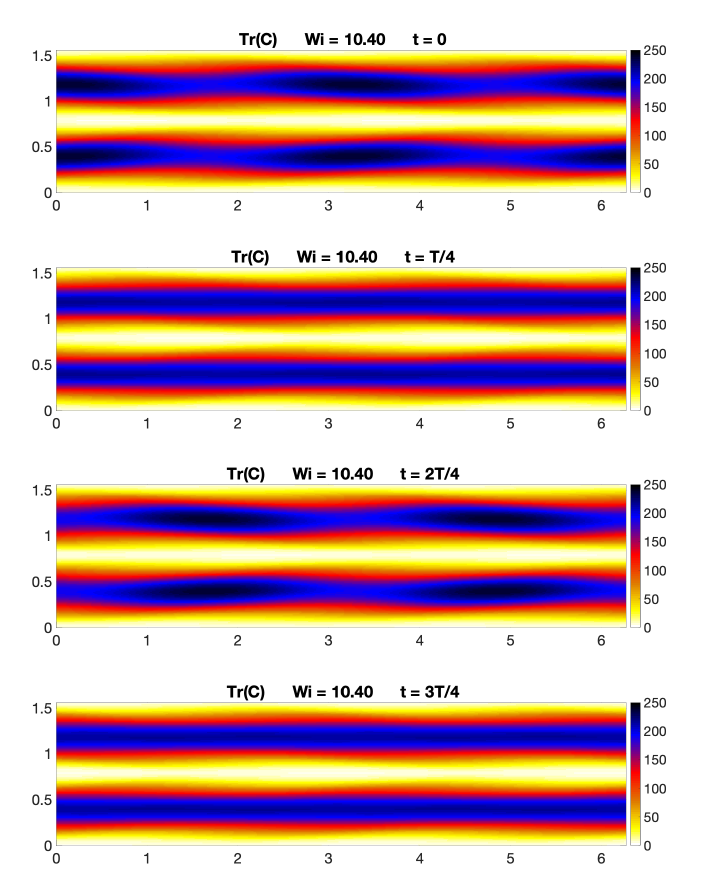}
    \caption{Standing waves over one period of oscillation for $\Wi=10.4$.  The period, {\sffamily{T}} $= 0.80$.}
    \label{fig:SticksPanels}
\end{figure}

Linear stability analysis (LSA) provides insight into the bifurcation where the analytic solution loses stability.  In addition to establishing a critical $\Wi$ (see Table~\ref{tab:Wi_c and Res}), we observe that two conjugate pairs of eigenvalues cross the imaginary axis together as $\Wi$ is increased, indicating a double Hopf bifurcation.  Just above the bifurcation, simulations produce standing waves in the polymer stress, which are discussed in Appendix \ref{A:standingwaves}.  The size of the imaginary part of the eigenvalues accurately predicts the frequency of oscillations of the standing waves.

For sufficiently large $\Wi$, LSA reveals that the analytic solution regains linear stability.  With $\nu=5\cdot10^{-4}$, this occurs at $\Wi_c=36.9.$  Both the upper and a lower critical $\Wi$ depend on the stress diffusion coefficient, $\nu$.  By varying $\nu$ we can construct a region of instability and gain insight into the effects of stress diffusion.  Further details are given in Appendix \ref{A:diff_stab}.

Just above the critical value $\Wi_c=9.61$ where the analytic solution becomes linearly unstable, there is a small range, observed in simulations from $9.65\leq \Wi \leq 11.2$, that supports solutions which are distinct from the others discussed in this work.  These solutions are standing waves in the strain energy density, as shown in Fig.~\ref{fig:SticksPanels}.  The waves alternate between two configurations ($t=0$ and $t=2${\sffamily{T}}$/4$ in the figure), each with four concentrated, slightly tilted ``stress islands.'' The difference between these two states is merely a horizontal shift by $\pi/2.$  Note, however, that the stress islands are \textit{not} traveling horizontally.  Rather, they disappear ($t=${\sffamily{T}}$/4$ and $t=3${\sffamily{T}}$/4$) and reform in the new location.

 \begin{figure}[h]
    \centering
    \includegraphics[width=0.9\linewidth]{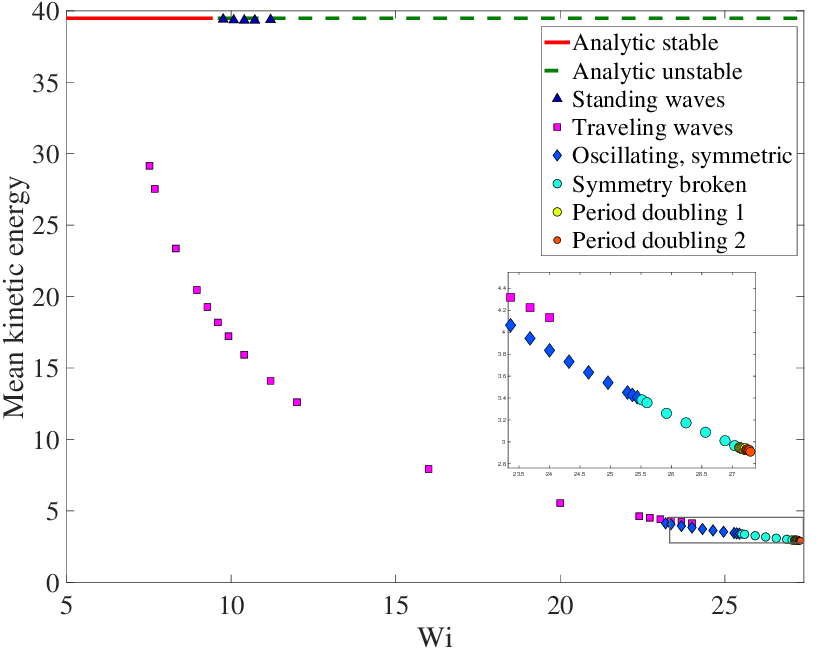}
    \caption{Mean kinetic energy and classification  of solution types as  a function of  $\Wi$. }
    \label{fig:bifdiag_ke}
\end{figure}

The standing waves emerge from the analytic solution in a continuous manner as $\Wi$ is increased, suggesting a supercritical bifurcation.  Continuing upwards to $\Wi>11.2$, traveling wave narwhals emerge, which are discussed in the main text. When the narwhals form, there is a large decrease in the kinetic energy, which is visible in the bifurcation diagram in Appendix \ref{A:KEbifdiag}.

\section{Kinetic Energy Bifurcation Diagram} \label{A:KEbifdiag}
The bifurcation diagram in the main text shows the strain energy as a function of $\Wi$.  In Fig.~\ref{fig:bifdiag_ke} we show the same bifurcation diagram in terms of the mean kinetic energy. The  decrease in energy in the subcritical bifurcation from analytic solutions to traveling wave solutions is evident when looking at the kinetic energy diagram. Note that the standing wave solutions have the higher level of kinetic energy, closer to the analytic solution.
     
\section{Symmetric Oscillating Solutions}\label{A:symmm}
\begin{figure}
    \centering
    \includegraphics[width=0.95\linewidth]{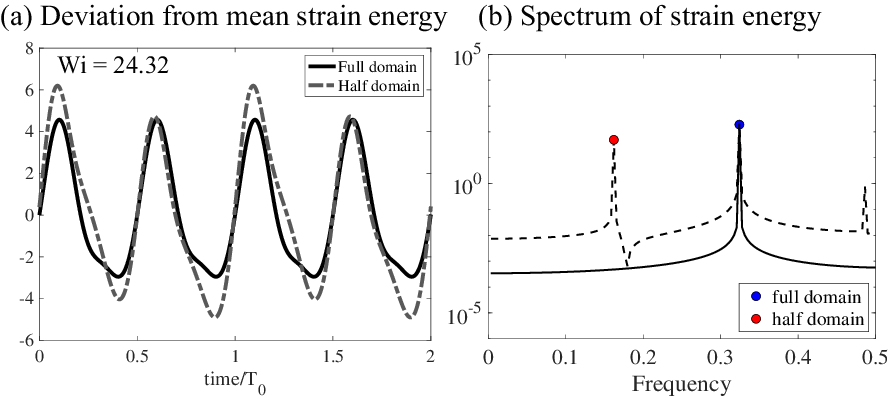}
    \caption{(a) The deviation from the mean strain energy averaged over both the full domain and bottom half of the domain. (b) The spectrum of the energy over  full and half domain show different dominant frequencies. The full domain has dominant frequency $0.3241$ (period 3.085),  and over half the domain the dominant frequency is $0.162$ (period 6.17).  This difference occurs due to the vertical symmetry for this solution.   }
    \label{fig:symmosc}
\end{figure}

In the main text it is noted that the first oscillatory solutions that emerge after traveling waves have a vertical symmetry which is later lost as $\Wi$ increases. This symmetry can be seen by examining the strain energy and spectrum over both the full domain as well as only the (bottom) half of the domain. Fig.~\ref{fig:symmosc} (a) shows deviations from the mean strain energy where the mean is taken over the full domain and also over only the bottom half of the domain.  When the period is computed using the dominant frequency over the full domain a period of 3.085 is calculated whereas over only the bottom half the period is 6.17, Fig.~\ref{fig:symmosc}(b). 

\section{Changes in stress distribution upon period doubling}\label{A:stresschanges}

\begin{figure}[h]
    \centering
    \includegraphics[width=0.95\linewidth]{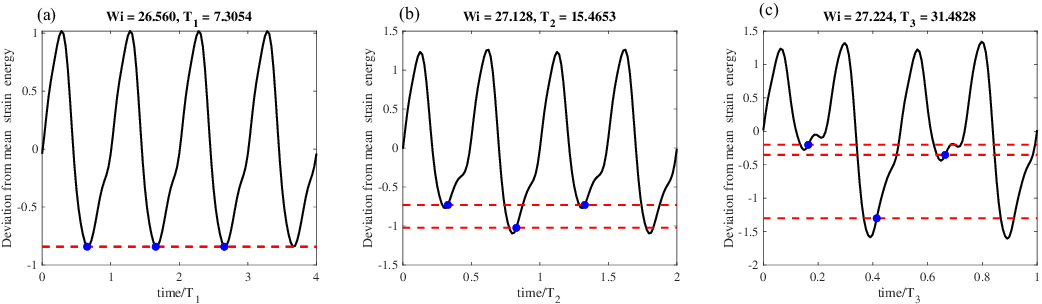}
    \caption{Deviation from mean strain energy as a function of normalized time for (a) $\Wi=26.56,$ period $T_1=7.3054,$ (b) $\Wi = 27.128,$ period $T_2 = 15.4653,$ and (c) $\Wi = 27.224,$ period $T_3 = 31.4828.$ The times highlighted in blue are separated by $T_1,T_2/2,$ and $T_3/4$ in (a)-(c) respectively.}
    \label{fig:panel_norms}
\end{figure}

As discussed in the manuscript,  when the traveling wave structure breaks down and oscillations arise, the dominant motion  is an up/down oscillation of  the narwhal  tusk. This motion has  period $\approx 7$ for all of the solutions in the period doubling cascade. When longer periods arise it is due to additional structures that are much more subtle to see in the stress distribution. To highlight these features we first  plot the deviation from the mean strain energy versus  time scaled by the period in Fig.~\ref{fig:panel_norms}.  We display simulations for $\Wi=26.56,27.128,$ and $27.224$ which have  period $T_1=7.3054, T_2 = 15,4653,$ and $T_3 = 31.4828,$ respectively. In  these plots we highlight 3 times with blue markers. The dotted red lines in the figures help illustrate that for $\Wi=26.56$ these times are separated by a period, and thus are at the same phase of the oscillation. For $\Wi=27.128$ the first and third time are separated by a period, while the second time is at a different phase of the oscillation, and for $\Wi=27.224$ these times correspond to three distinct phases.  Note that in all cases these specific times are separated by $\approx 7$, hence these are all in the same phase of the dominant up/down motion of the tusk.

\begin{figure}
    \centering
    \includegraphics[width=0.95\linewidth]{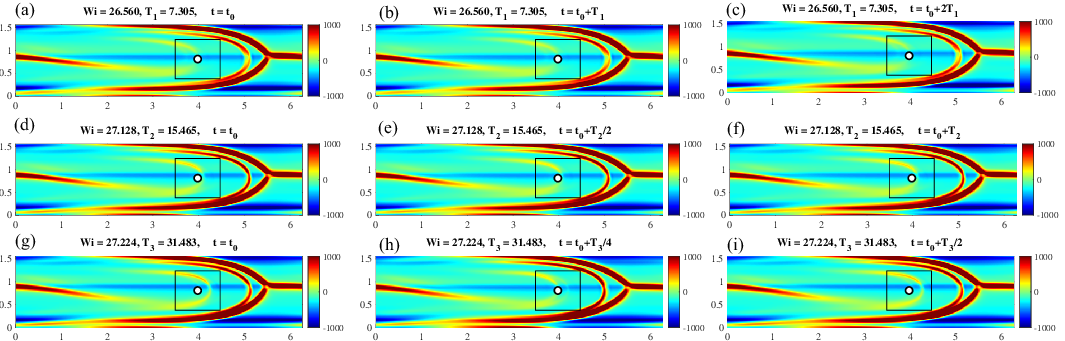}
    \caption{The colorfields show deviations in $\trC$ from the mean strain energy  for the simulations at (a)-(c) $\Wi=26.56$, (d)-(f) $\Wi = 27.128$, and (g)-(i) $\Wi = 27.224.$ The times correspond to the times highlighted in Fig.~\ref{fig:panel_norms}.  The white marker is at the same location in each panel to highlight subtle changes near this point, and the rectangle indicates the region displayed in the zoom in Fig.~\ref{fig:panel_figz}. A constant approximate traveling wave speed has been removed so that the tusk is in nearly the same location at all times.}
    \label{fig:panel_fig}
\end{figure}

\begin{figure}
    \centering
    \includegraphics[width=0.95\linewidth]{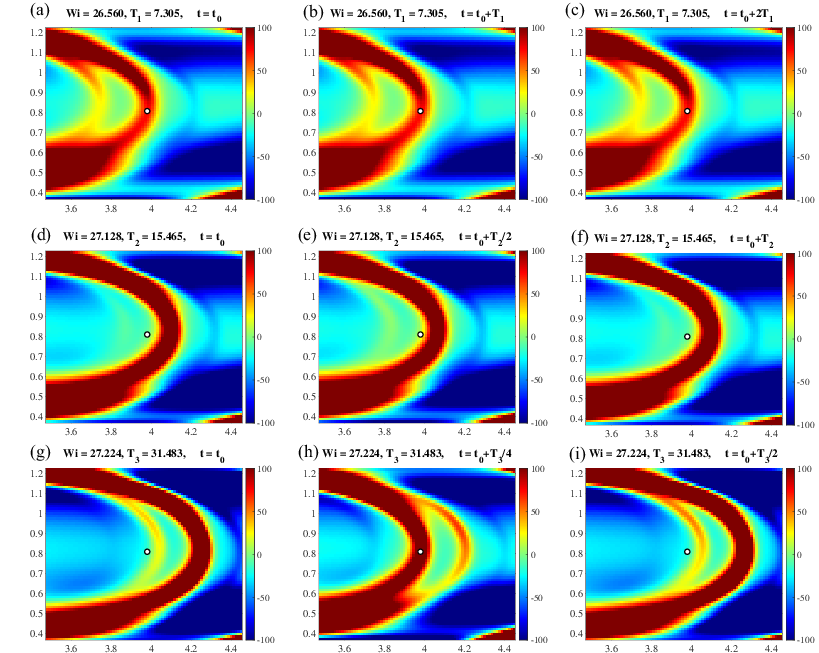}
    \caption{Spatial zoom of colorfields for the deviation of $\trC$ from the mean of $\trC$ in the rectangular regions from Fig.~\ref{fig:panel_fig}. Subtracting the mean and using a tighter range of the colorbar allows for more subtle changes to be visualized. The white marker is at the same location in each panel.}
    \label{fig:panel_figz}
\end{figure}

In Fig.~\ref{fig:panel_fig} we show colorfields of deviations in $\trC$ from the mean strain energy for the simulations at  (a)-(c) $\Wi=26.56$, (d)-(f)$\Wi = 27.128$, and (g)-(i) $\Wi = 27.224$ sequentially at the three times highlighted in Fig.~\ref{fig:panel_norms}.  The white marker is at the same location in each panel to highlight subtle changes near this point, and in 
  Fig.~\ref{fig:panel_figz} we plot a zoom of the region highlighted by the rectangles. 
Although these  are not strictly traveling wave solutions we compute an approximate traveling wave speed and shift the frames into these approximate traveling wave coordinates to compare images with the tusk in nearly the same location at all times. 

 The changes that occur upon period doubling are subtle and most obvious in Fig.~\ref{fig:panel_figz}.  For $\Wi = 27.128,$ panels (d) and (e) correspond to different phases in the oscillation and the large stress band to the right of the marker comes a bit closer to the marker between $t_0$ and $t_0+T_2/2.$  For $\Wi = 27.224,$ panels (g),(h) and (i) are all at different phases, and it is more clear that the large stress band is moving horizontally. There are much more subtle changes between panels (g) and (i), reminiscent of the subtle changes between (d) and (e). Looking back at Fig.~\ref{fig:panel_norms}, the differences in the energy are also largest for the times corresponding to those from (g) and (h).

\section{Diffusion} \label{A:diffusion}

\subsection{Lower Diffusion Results} \label{A:LowDiffResults}

Upon finding the route to chaos presented in the main text, a natural question to ask is whether the size of the stress diffusion (with $\nu=5\cdot10^{-4}$) played an important role in the dynamics.  To help answer this question, we ran a select number of simulations with $\nu=5\cdot10^{-5}$.  This is comparable to the value determined in \cite{MorozovCoherentStructures} to realistically arise from kinetic theory.  With this lower value of diffusion, we found that the route to chaos via period doubling bifurcations remains intact, and discovered no indications of substantially different dynamics as a result of reducing diffusion. The low diffusion simulations revealed traveling waves and oscillations which first break symmetry and then undergo a period doubling.  Simulations with this amount of diffusion required a grid resolution of $1024\times256$ and a timestep of $\Delta t=6.25\cdot10^{-4}$.

\begin{figure}[H]
    \centering
    \includegraphics[width=0.75\textwidth]{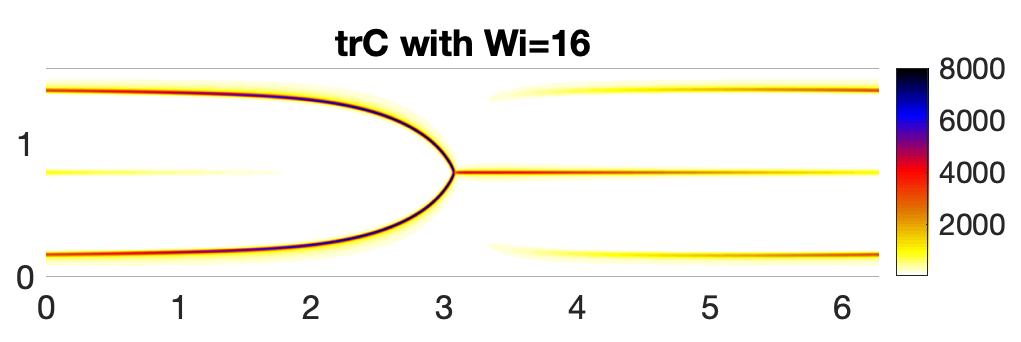}
    \caption{Strain energy density of a traveling wave solution at $\Wi=16$ with $\nu=5\cdot10^{-5}$.}
    \label{fig:low diff TWS}
\end{figure}

From $8.8\leq\Wi\leq22.4$, we obtained traveling wave solutions taking the form of narwhals.  An example is shown in Fig.~\ref{fig:low diff TWS}.  These contain exceptionally high peak stresses with sharp gradients, which necessitated a smaller timestep ($\Delta t=3.125\cdot10^{-4})$ than other simulations to maintain stability.

Oscillatory narwhals emerged at higher $\Wi$ through a subcritical bifurcation.  The oscillating behavior was continued downwards to $\Wi=21.6$, demonstrating the coexistence of oscillatory narwhals with steady TWS narwhals.  The first oscillations to emerge are characterized by a vertical symmetry which results in the kinetic and strain energies oscillating at half the period of the narwhal's oscillation.  The energies' oscillatory periods in this regime range from $2.68$ to $3.12$.

\begin{figure}[H]
    \centering
    \includegraphics[width=0.4\linewidth]{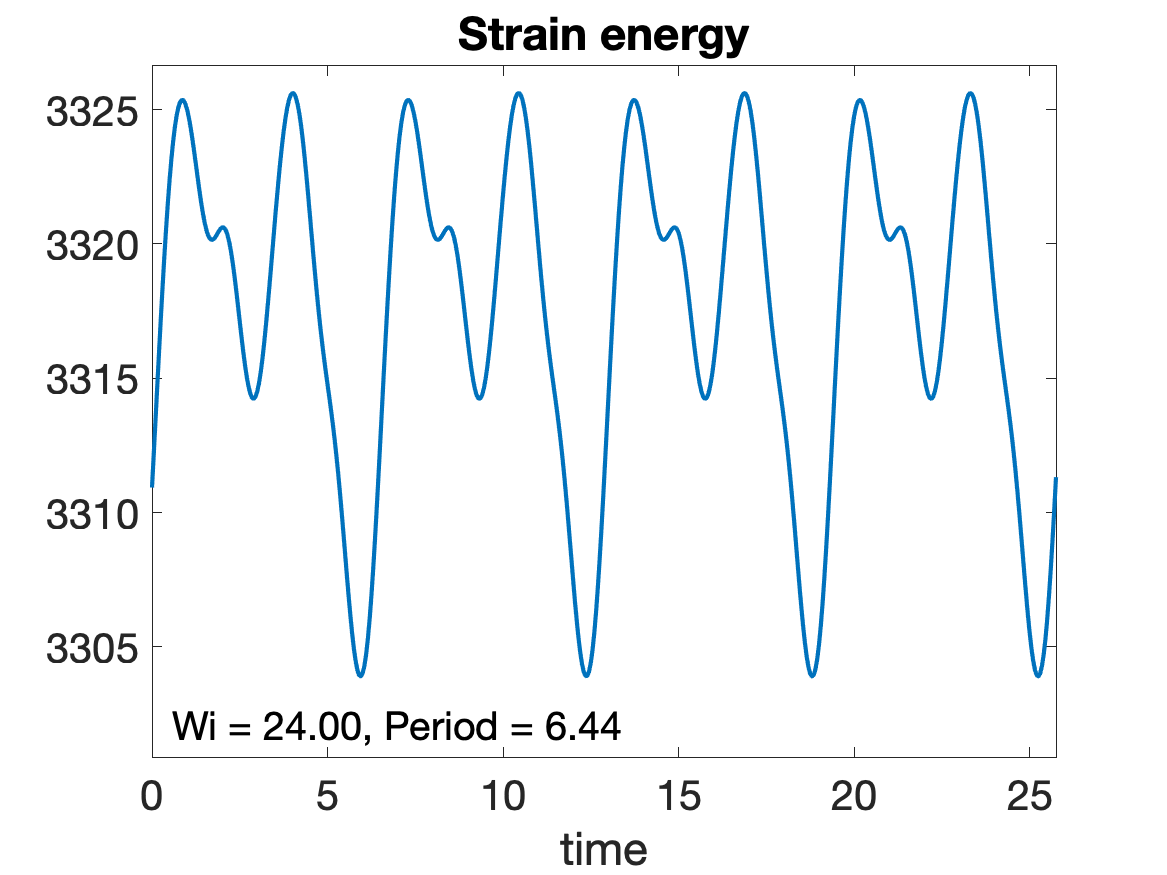} \includegraphics[width=0.4\linewidth]{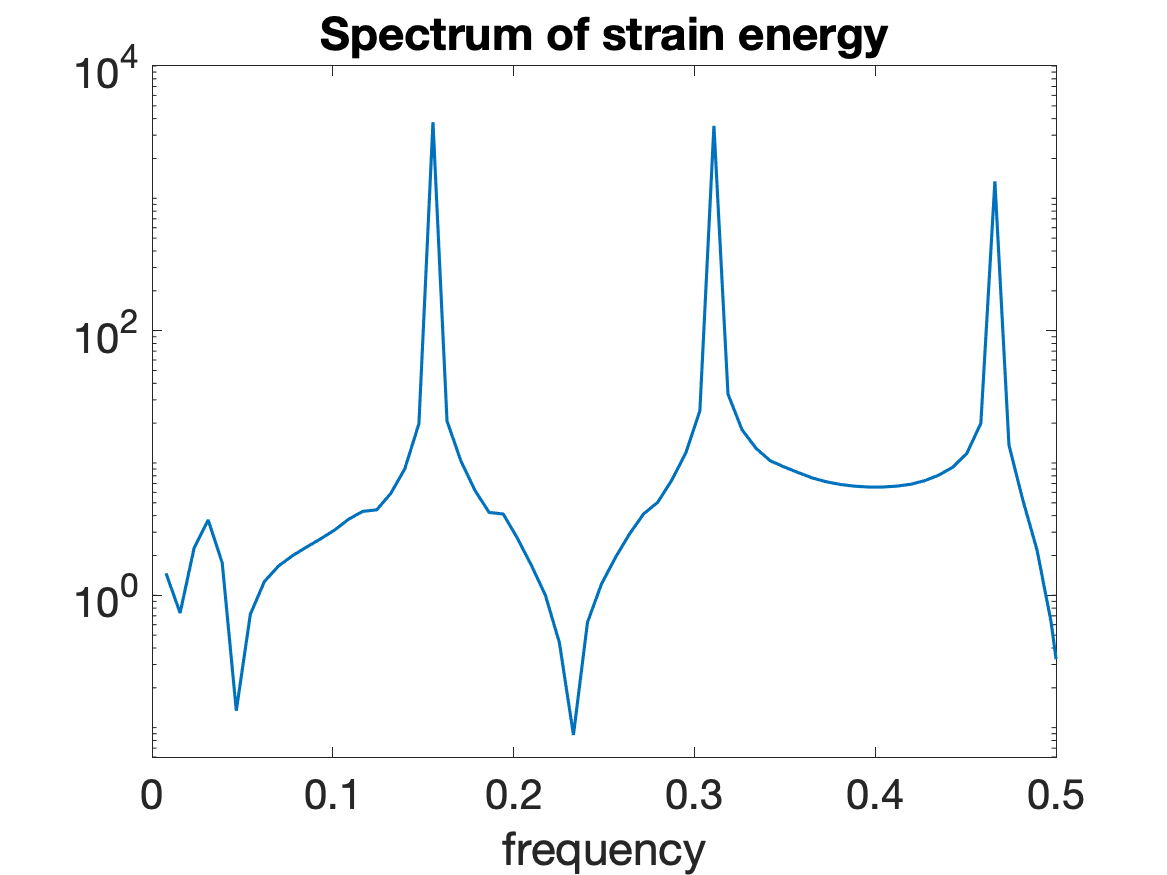}
    \includegraphics[width=0.4\linewidth]{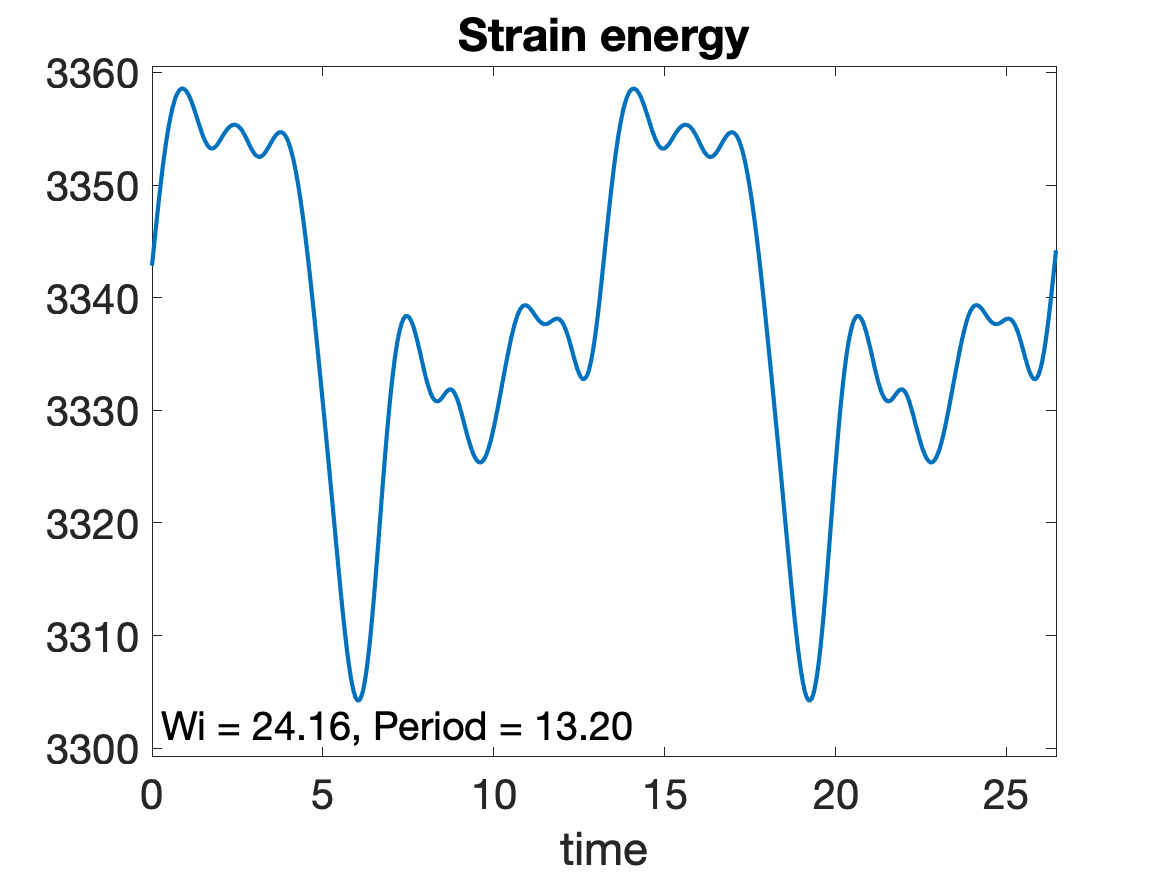} \includegraphics[width=0.4\linewidth]{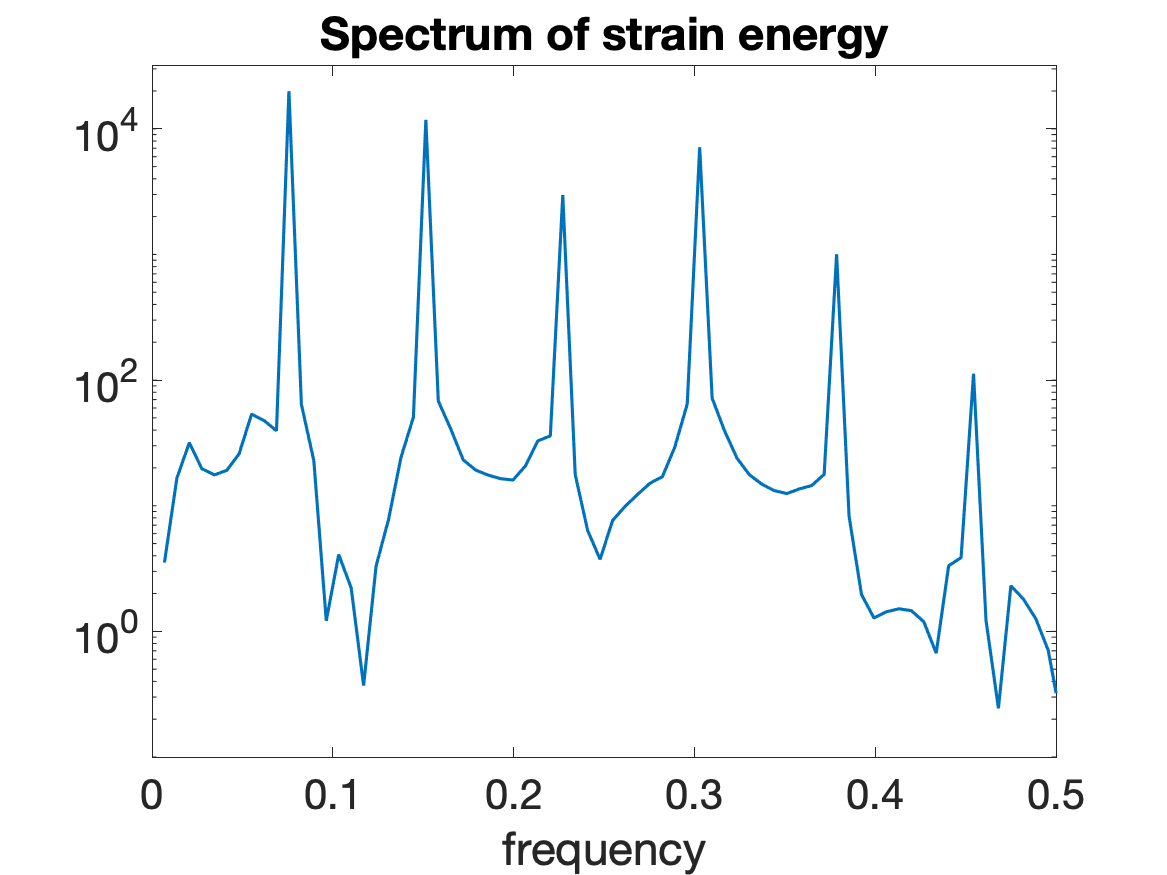}
    \caption{A period doubling with $\nu=5\cdot10^{-5}$ is apparent in the strain energy over time (left column) and the spectra of the strain energies (right column).}
    \label{fig:DoublingLowDiff}
\end{figure}

Continuing upwards in $\Wi$, the symmetry in the oscillations breaks, and at $\Wi=24$ the energies oscillate with a period of $6.44$.  Above this, a period doubling occurs and at $\Wi=24.16$ the period is $13.20$.  The strain energy over time and its spectrum before and after this period doubling is shown in Fig.~\ref{fig:DoublingLowDiff}.

\subsection{Stability Region} \label{A:diff_stab}

\begin{figure}[h]
    \centering
    \includegraphics[width=0.75\linewidth]{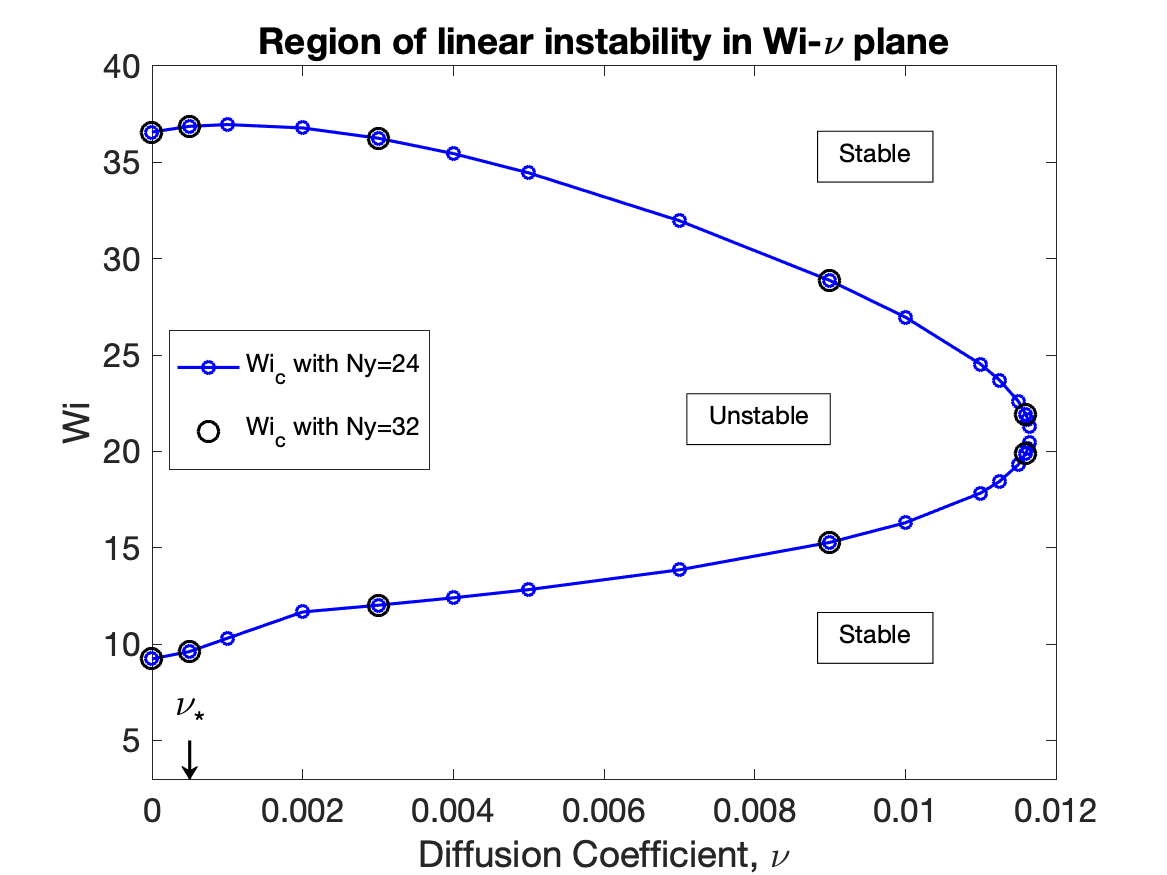}
    \caption{Region of instability for viscoelastic Kolmogorov flow.  The amount of diffusion used for the results in the main text is marked on the horizontal axis, $\nu_*=5\cdot10^{-4}$}
    \label{fig:diffusionAndWiC}
\end{figure}

As mentioned in Appendix \ref{A:LinStabResults}, linear stability analysis reveals that as $\Wi$ is increased, viscoelastic Kolmogorov flow first loses stability and later regains it.  The critical values of $\Wi$ where this occurs depend on the amount of stress diffusion.  In Fig.~\ref{fig:diffusionAndWiC} we report the region in the $\Wi-\nu $ plane where the flow is linearly unstable.  Diffusion is generally stabilizing (the exception being small amounts of diffusion resulting in an increase in the upper critical $\Wi$), and above $\nu \approx 1.17\cdot 10^{-2}$, the instability disappears entirely.

\section{Checking Resolution} \label{A:checkres}
We ran a select number of simulations at the original size of diffusion, $\nu = 5 \cdot 10^{-4}$, and twice the resolution. That is, on a $1024\times256$ grid. Our objective was to verify the values of Wi around which period doubling bifurcations occur.
					
Testing began with the end states of simulations at our standard resolution (512 $\times$ 128), with $\Wi$ near bifurcation points. The resolution was then doubled and the simulations continued. In all cases, we found the expected periods of oscillations which support our conclusions.  These tests are summarized in Table~\ref{HiResTable}

\begin{table}
    \centering
    \begin{tabular}{|c|c|c|c|}
        \hline
        ${\Wi}$ & {Simulation Length} & \shortstack{Period at \\ standard resolution}  & \shortstack{Period at \\ high resolution}\\
        \hline
        27.04  & 370 & 7.66 & 7.66 \\
        \hline
        27.12  & 380 & 7.72 & 7.72 \\
        \hline
        27.128 & 500 & 15.47 & 15.46 \\
        \hline
        27.216 & 500 & 15.73 & 15.72 \\
        \hline
        27.224 & 500 & 31.48 & 31.46 \\
        \hline
        27.296 & 500 & 31.74 & 31.73 \\
        \hline
    \end{tabular} 
    \caption{Verification of $\Wi$ values around period doublings with high resolution simulations.}
    \label{HiResTable}
\end{table} 

Next, we confirmed that the bifurcations would still occur after continuation at higher resolution. We used initial data from a simulation below the bifurcation and doubled the resolution. Then we increased the Weissenberg number beyond the bifurcation, added a small perturbation and continued the simulations. These tests were performed around each bifurcation point and around the $\Wi$ value where the solution breaks symmetry. In each case the solution approached a periodic solution with the expected period.

\section{Movies}\label{A:movies}

Several movies are provided at the following url \url{https://drive.google.com/drive/folders/1BMI8jiEhaTRZxfKvr_ZhW2Wfx6-V9HiI?usp=drive_link}.  Some include passively advected markers and some are placed in traveling wave coordinates (TWC). Movie descriptions and filenames can be found in Table~\ref{tab:movieTab}.


\begin{table}[htb]
\includegraphics[clip, trim=2cm 8.5cm 1cm 5cm, width=.7\textwidth]{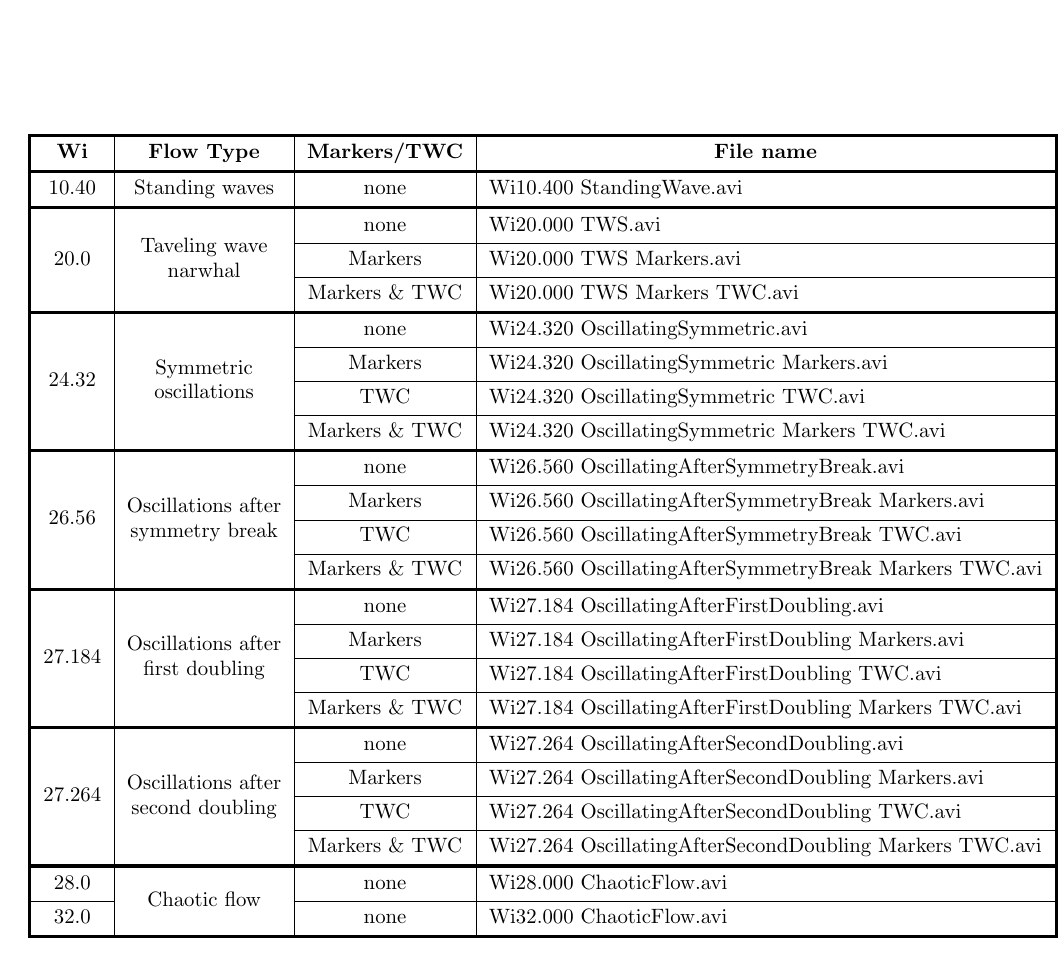}
\caption{Movie descriptions and filenames} 
\label{tab:movieTab} 
\end{table}

\newpage

\bibliography{bib_arxiv}

\end{document}